\documentclass[preprint2]{aastex}
\shorttitle{ 
'SZ Observations of Strong Lensing Clusters'  } 
\shortauthors{ 
Gralla et al.}
\usepackage{graphicx}
\usepackage{amssymb}
\usepackage{epstopdf}
\usepackage{epsfig}

\begin{document}

\title{Sunyaev Zel'dovich Effect Observations of Strong Lensing Galaxy Clusters: Probing the Over-Concentration Problem}
\author{Megan B. Gralla\altaffilmark{1,2},
Keren Sharon\altaffilmark{2},
Michael D. Gladders\altaffilmark{1,2},
Daniel P. Marrone\altaffilmark{1,2,3},
L. Felipe Barrientos\altaffilmark{4}
Matthew Bayliss\altaffilmark{1,2},
Massimiliano Bonamente\altaffilmark{5},
Esra Bulbul\altaffilmark{5},
John E. Carlstrom\altaffilmark{1,2,6,7},
Thomas Culverhouse\altaffilmark{8},
David G. Gilbank\altaffilmark{9},
Christopher Greer\altaffilmark{1,2},
Nicole Hasler\altaffilmark{5},
David Hawkins\altaffilmark{8},
Ryan Hennessy\altaffilmark{1,2},
Marshall Joy\altaffilmark{10},
Benjamin Koester\altaffilmark{1},
James Lamb\altaffilmark{8},
Erik Leitch\altaffilmark{1,2},
Amber Miller\altaffilmark{11,12},
Tony Mroczkowski\altaffilmark{13,14},
Stephen Muchovej\altaffilmark{8},
Masamune Oguri\altaffilmark{15,16},
Tom Plagge\altaffilmark{1,2},
Clem Pryke\altaffilmark{1,2,6},
David Woody\altaffilmark{8}
}

\altaffiltext{1}{Department of Astronomy \& Astrophysics, University of Chicago, Chicago, IL 60637}
\altaffiltext{2}{Kavli Institute for Cosmological Physics, University of Chicago, Chicago, IL 60637}
\altaffiltext{3}{Hubble Fellow}
\altaffiltext{4}{Departamento de Astronom\'ia y Astrof\'isica, Pontificia Universidad Cat\'olica de Chile, 7820436 Macul, Santiago, Chile}
\altaffiltext{5}{Department of Physics, University of Alabama, Huntsville, AL 35812}
\altaffiltext{6}{Enrico Fermi Institute, University of Chicago, Chicago, IL 60637}
\altaffiltext{7}{Department of Physics, University of Chicago, Chicago, IL 60637}
\altaffiltext{8}{Owens Valley Radio Observatory, California Institute of Technology, Big Pine, CA 93513}
\altaffiltext{9}{Department of Physics and Astronomy, University of Waterloo, Waterloo, Ontario, Canada N2L 3G1}
\altaffiltext{10}{Space Science Office, VP62, NASA/Marshall Space Flight Center, Huntsville, AL 35812}
\altaffiltext{11}{Columbia Astrophysics Laboratory, Columbia University, New York, NY 10027}
\altaffiltext{12}{Department of Physics, Columbia University, New York, NY 10027}
\altaffiltext{13}{Department of Physics and Astronomy, University of Pennsylvania, Philadelphia, PA 19104}
\altaffiltext{14}{Einstein Fellow}
\altaffiltext{15}{Kavli Institute for Particle Astrophysics and Cosmology, Stanford University, 2575 Sand Hill Road, Menlo Park, CA 94025, USA}
\altaffiltext{16}{Division of Theoretical Astronomy, National Astronomical Observatory of Japan, 2-21-1 Osawa, Mitaka, Tokyo 181-8588, Japan}

\begin{abstract}
We have measured the Sunyaev Zel'dovich (SZ) effect for a sample of ten strong lensing selected galaxy clusters using the Sunyaev Zel'dovich Array (SZA).
The SZA is sensitive to structures on spatial scales of a few arcminutes,
while the strong lensing mass modeling constrains the mass at small scales (typically $<30\arcsec$). Combining the two 
provides information about the projected concentrations of the strong lensing 
clusters.  
The Einstein radii we measure are twice as large as expected given the masses
inferred from SZ scaling relations.
A Monte Carlo simulation indicates that a sample randomly drawn from the 
expected distribution would have a larger median Einstein radius than the observed clusters about 3\% of the time.
The implied overconcentration has been noted in previous studies with smaller
samples of lensing clusters. It persists for this sample, with the caveat that this could result from a systematic
effect such as if the gas fractions of the strong lensing clusters are substantially below what is expected.
\end{abstract}

\section{Introduction}

Earlier studies have found that strong lensing galaxy clusters are more concentrated than simulations predict 
\citep[e.g., ][]{broadhursta1689, broadhurstovercon, ogurifirst, hennawi}.  One possible explanation is that dark
matter halos form at earlier times than the standard $\Lambda$CDM expectation \citep[e.g., ][]{wechsler}, challenging this otherwise well-supported 
cosmological model.
Exotic physics such as non-Gaussianity or early dark energy could cause such an early collapse of dark matter halos \citep[e.g., ][]{mathis, fedeli}.
Another explanation for the observed over-concentration could be that current models do not properly incorporate the effects of baryons 
on the core density profiles in clusters \citep[e.g., ][]{eduardo}.  The discrepancy could also result from inaccuracies in current cluster models 
of strong lenses due to extrapolations of the concentration probability distribution from more easily simulated, lower mass systems 
\citep[discussed in ][]{oguri}.

\citet{broadhurst} recast this overconcentration problem into a comparison of the cluster mass at small scales with the cluster mass at large scales.
They measured the inner mass through strong lensing mass modeling and expressed it in terms of the Einstein radius, $\theta_{E}$, which for a 
circularly symmetric lens with an arbitrary mass profile satisfies $\theta_{E}^{2} \propto M(<\theta_{E})$ \citep{lensreference}.
They measured the overall cluster mass through weak lensing.  This approach is useful because it enables a comparison to theoretical predictions
without measuring the concentration of the cluster profile, which would require knowledge of the shape of the density profile over 
a wide range of scales in the cluster, some of which are not well constrained by the available data.  For a sample of four strong lensing clusters,
they found mass measurements that are discrepant with theoretical predictions and interpreted this disagreement as a conflict between
their observations and $\Lambda$CDM cosmology. 
Other authors have constructed strong lensing models for larger samples of massive clusters, often selected on the basis of other mass proxies,
such as X-ray luminosity, rather than as strong lenses.  These studies also find that the distribution of measured Einstein radii are
larger than expected, although the significance of the discrepancy is small \citep{locusswl1, macswl}.

Using the Sunyaev-Zel'dovich Array (SZA), we have observed the Sunyaev Zel'dovich (SZ) effect for a sample of ten clusters selected from large optically
selected cluster catalogs for 
their strong lensing signatures.  We derive mass measurements from these SZ observations and Einstein radii measurements from detailed
strong lensing mass modeling, comparing our results with theoretical predictions. 
A strength of using ICM-based mass estimates for strong lensing clusters is that they have less scatter at fixed mass than optical richness measurements and 
different dependencies on 
line-of-sight structure than weak lensing measurements (strong lenses are expected to be preferentially aligned along the line of sight).
This paper is part of a larger program to obtain a variety of mass estimates for strong lensing clusters (weak lensing, velocity dispersion, X-ray).

Throughout this work we assume a flat $\Lambda$CDM cosmology with $\Omega_{m} = 0.27$, $\Omega_{\Lambda} = 0.73$, and $h = 0.73$ in order
to calculate the angular diameter distances to the clusters and sources.  The theoretical models are derived adopting the WMAP5 cosmology \citep{wmap5}
as the fiducial cosmological model.

\section{Data and Analysis}

\subsection{Sample selection}

We select strong lensing clusters from two programs: the Sloan Giant Arc Survey (SGAS) and the Red-Sequence 
Cluster Survey Giant Arc program (RCS-GA).  
The selection of strong lensing clusters for SZA observations was driven by RA constraints and
availability of complementary data (primarily spectroscopy to facilitate
lens modeling).

SGAS is a survey to detect strong lensing by clusters with the clusters
selected via the red-sequence method \citep{redsequence} applied to the
Sloan Digital Sky Survey (SDSS) DR7 photometry. Lenses have been selected in
two ways: either by imaging of rich clusters to search for lensing features
\citep{sgas} or from clusters and groups (not necessarily rich)
that show evidence for lensing directly in the SDSS imaging \citep{sloanarcs}.
Candidates were subsequently confirmed by further imaging. In either
case, deeper imaging was obtained at the Wisconsin-Indiana-Yale NOAO 3.5m
telescope, the University of Hawaii 88 inch telescope, or the 2.5m Nordic
Optical Telescopes. SGAS targets in this paper are drawn from both samples.

The RCS-GA is a program to detect strong lensing around clusters in the second Red-Sequence Cluster Survey\footnote[1]{http://rcs2.org} 
\citep[RCS2; ][]{rcs2}.  
RCS2 clusters are also selected photometrically using the
red-sequence method.  The survey covers nearly 1,000 square degrees, with imaging in \emph{g'}, \emph{r'}, \emph{z'} filters from the Canada-France-Hawaii 
Telescope.  The typical $5\sigma$ point source depth is 25.3 magnitudes in the \emph{g'} band, 24.8 magnitudes in the \emph{r'} band, and 22.5 magnitudes
in the \emph{z'} band.  Strong lensing features were identified by eye from 3-color composite images of clusters directly from the survey data.

Because arc redshifts are important for constraining strong lensing mass models, we only selected strong lenses with 
spectroscopically measured arc redshifts.
We supplemented our sample with two additional clusters from the literature,
one from the first Red-Sequence Cluster Survey (RCS1), RCS1J2319, and a well-known strong lensing Abell cluster, 
Abell 1689.  Table \ref{sample} contains information about the sample.  

\subsection{SZ Observations and Analysis}\label{sz}

The SZA is an 8-element interferometer originally located at the Owens Valley Radio Observatory in California that is now incorporated into the
Combined Array for Research in Millimeter-wave Astronomy (CARMA) at a nearby 2200\,m elevation site at Cedar Flat in the Inyo 
Mountains of California.  
All of the observations were conducted at
a center frequency of 31 GHz, for which the SZA has baselines of 350$-$1300$\lambda$, corresponding to spatial scales
\footnote{This refers to the wavelength of the Fourier component measured; the largest features imaged well in a map are about a factor of two smaller.} of 9.8\arcmin\ to 2.6\arcmin\ for 
sensitivity to the cluster SZ effect and baselines of 2$-$7 k$\lambda$ for the simultaneous measurement of compact radio sources.  Radio sources were found within 
1 arcminute of the target center for 7 out of the 10 targets.  
The on-source, unflagged observing
time for each cluster ranged from 11 to 57 hours and resulted in rms noise of 0.23 to 0.08 mJy/beam for the short baselines.
\cite{sza} provide a detailed description of the SZA instrument, measurements and the data reduction pipeline developed by the SZA 
collaboration. 

We use Markov Chain Monte Carlo (MCMC) analysis software \citep{markov} to fit a model for the gas pressure to
the visibility data.  The model is based on the \citet{nagai} simulations with parameters fit to X-ray data.  The pressure profile is given by 
\begin{equation} 
P(r) \propto x^{-\gamma} (1 + x^{\alpha})^{-(\beta-\gamma)/\alpha}
\end{equation}
 where $x \equiv r/r_{s}$, where $r_{s}$ is the scale radius of the profile, and with parameters $\alpha$, $\beta$, and $\gamma$ set to 0.9, 5.0, and 0.4 
respectively.  \citet{tony} found that this model fits both SZ and X-ray data well.
For each cluster, we determine the best fit parameters for the pressure profile normalization and scale radius and used these 
in all subsequent analyses.  The fluxes of compact radio sources were fit along with the cluster parameters.

The observable from the SZ effect is the integrated Compton \emph{y}-parameter, Y, which is proportional to the integrated pressure of the 
intracluster gas.  
We calculate a spherically integrated SZ parameter, Y$_{500,sph}$, by integrating the pressure profile to $r_{500}$, defined as the radius within which 
the enclosed mass is equal to 500
times the critical density of the universe.  We chose this radius to use available scaling relations.

Masses are calculated from the integrated Y; we used a scaling relation based on SZA observations and weak lensing measurements of clusters in the 
Local Cluster Substructure Survey (LoCuSS) sample.  Details of the LoCuSS sample will be found in \citet{graham}, and details of the 
weak lensing analysis are found in \citet{okabe}.  The LoCuSS SZ measurements and analysis  will be 
published in an upcoming paper \citep{marronetobe}.  The LoCuSS clusters span a similar range in integrated Y (when scaled by E(z)$^{-2/3}$, as expected
for self-similar scaling) as the strong lensing clusters, 
although they lie at lower redshifts.   We fixed the slope of the scaling relation to the self-similar value of 0.6.
Lacking a priori knowledge of the radius r$_{500}$, we iteratively solve for r$_{500}$ and M$_{500}$ through the scaling relation;
for each iteration we integrate the pressure profile to the r$_{500}$ from the previous iteration to calculate Y$_{sph}$, 
then calculate M$_{500}$ from the scaling relation, calculate r$_{500}$ according to this M$_{500}$, and repeat until this converges. 
See Table \ref{sample} for the resulting r$_{500}$, Y$_{sph,500}$ and M$_{500}$ for each cluster in this sample.  

\subsection{Lensing mass models}

Einstein radii for the clusters in this sample were computed from strong-lensing models that were either derived for this work or taken from
the literature (Table~\ref{sample}). When taken from the literature, the models were adjusted according to our fiducial cosmology.
All the lensing models that are utilized here were constructed using the publicly-available software
{\tt lenstool} \citep{lenstool}, with MCMC minimization in the source plane. A detailed description of the lens modeling process will be
published elsewhere \citep{lensingmodelsinprep}. In short, we consider only clusters with secure spectroscopic redshifts for both the cluster
and at least one source. The redshifts and positions of multiply-imaged arc systems (sources) are used as constraints, and
when available, cluster velocity dispersions and low-uncertainty photometric redshifts of secondary arcs are used as priors. The clusters
are typically modeled with several mass halos, depending on the complexity that is needed: a \citet[][(NFW)]{nfw} or Pseudo-Isothermal Ellipsoid Mass
Distribution (PIEMD) to represent the cluster mass; a secondary PIEMD to represent a group-scale halo or external shear; and contributions
from cluster-member galaxies, represented by PIEMDs, with parameters that follow their observed values and scale with luminosity \citep[see ][]{limousina1689}.

The model-inferred Einstein radius is defined as $\theta_{E}=\sqrt{A}/\pi $, where $A$ is the area inside the tangential (outer) critical curve.
For an elliptical critical curve, this is equivalent to the definition that \cite{oguri} used; we have simply generalized it to more
complex lensing models.

To compare the observed Einstein radii for clusters of different redshifts to theoretical predictions, we 
re-compute $\theta_{E}$ of each cluster as follows. For each
cluster, we derive a lensing mass reconstruction as explained above, according to the cluster and measured arc redshifts. We then use the derived
mass distribution, assign to it the fiducial cluster redshift of 0.5 (near the average cluster redshift of the sample), 
and compute $\theta_{E}$ for a source plane at $z=2.0$. The resulting Einstein radii
for the actual cluster and source redshifts as well as for the fiducial ones are listed in Table~\ref{sample}. 

The uncertainties of the model-dependent Einstein radii are estimated through simulation. For each cluster, we compute $\theta_{E}$ in a series of
lens models, each one with a set of parameters drawn from steps in the MCMC that are within [$\chi^2_{min}$,$\chi^2_{min}+2$]. For most of
the clusters, the simulated $\theta_{E}$ values differ from that of the best-fit model by less than $5\%$, although there are a few notable 
exceptions.
RCS1J2319 and SDSS2111-01 are only constrained by arcs that lie on one side of the
cluster center.
This means that although the critical curves in all the accepted models are tightly constrained to pass through these arcs, a variety of
critical curve shapes and sizes are allowed by the data, resulting in larger uncertainties on the $\theta_{E}$ values (up to $\sim40\%$).  
In the case of Abell 963, the cluster mass
is degenerate with the position of the cluster centroid, resulting in a similarly large uncertainty in $\theta_{E}$ ($33\%$).


As noted above, the location of the critical curve is directly determined by observations.  If we were to use in our analysis $\theta_{E}$ as derived
for the actual cluster and arc redshifts, the uncertainties estimated above would be satisfactory. However, for some clusters the
extrapolation of the models to the fiducial redshift may result in an additional uncertainty.  Sharon et al. (2010, in prep) show that
the total projected mass inside the critical curve can be measured to within a few percent, even with a limited number of multiply-lensed
images -- as is the case with most of the clusters in this sample. However, outside the range of radii in which arcs are observed, the
enclosed mass becomes uncertain. Assuming that $M(<\theta_{E}) \propto \theta_{E}^{2}$ \citep{lensreference} and thus $\Delta M(<\theta_{E})/M(<\theta_{E}) \propto \Delta \theta_{E}/\theta_{E}$,
we argue that the accuracy with which $\theta_{E}$ is known is only as good as the accuracy of the enclosed mass.  

We therefore examine each $R_E$ with respect to observed projected
distances of the arcs in each cluster. In all but three clusters, the
fiducial $R_E$ are within the radial range covered by observed arcs.
For these clusters we conservatively increase the relative uncertainty
of $R_E$ by 5\%. For the other three clusters, Abell 963, SDSS J1343,
and SDSS J1531, we increase the relative uncertainty by 5.5\%, 5.5\% 
and 8.9\%, respectively, based on the relative scatter in mass
measurement as a function of radius from \citet{keren2}.

\subsection{Theoretical predictions} \label{theory}

\citet{oguri} calculated the expected distributions of large Einstein radii for triaxial dark matter halos based on a semi-analytic model.
They randomly generated a catalog of massive dark matter halos according to a mass function, assigning each halo 
a density profile following a triaxial generalization of the NFW profile, with the axis ratio and concentration parameter randomly
assigned according to the probability distributions determined by {N}-body simulations of dark matter halos in \citet{js02}.  
They projected these triaxial halos in two dimensions and calculated
the resulting projected convergence and shear maps.  Einstein radii were defined as the geometric mean of the distances from the halo center
to the outer critical curve along the major and minor axes of the projected two-dimensional density distribution.  They produced 300 
realizations of these mock all-sky catalogs of Einstein radii.  For further details, see \citet{oguri}.

Based on these models, we generate an average relation between the Einstein radii and M$_{500}$ for a large Einstein radius sample
cluster population for the same fiducial lens and source redshifts that are used to calculate the Einstein radii (0.5 and 2.0, respectively). 
This effectively is a measure of the projected concentration, but using the two observables corresponding 
to the two different measurements we made for our real sample of clusters. 

A strong lensing selected sample tends to have larger Einstein radii than a sample selected based on other mass observables.  We have
included this selection effect in the predictions for the relation between the Einstein radii and M$_{500}$ by weighing the
probability distribution with the square of $\theta_{E}$.
We note that the average Einstein radius is not noticeably affected by including a brightness cut on the arcs that mimics our selection.

\section{Results and Discussion}

\subsection{Mass and Einstein Radius}

\begin{figure}
\plotone{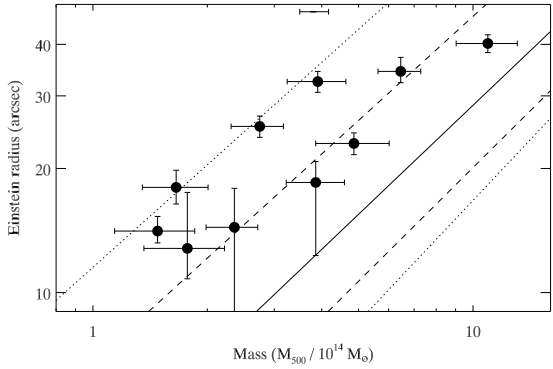}
\caption{The Einstein radii versus M$_{500}$ (as derived from SZ measurements) for strong lensing clusters. The solid line shows the 
median of the theoretical halo models
that have been selected to be efficient strong lenses.  The dashed lines enclose 68$\%$ of such models, and the dotted lines enclose 90$\%$.
 All of the lensing models have been scaled to a fiducial lens redshift of 0.5 and 
a fiducial source redshift of 2.0.  The error bar at the top of the plot shows the median uncertainty in mass from the uncertainty in 
the normalization of the scaling relation used to determine the masses from the SZ observations.\label{mvr}}
\end{figure}

The main result of this letter is shown in Figure \ref{mvr}.  This figure shows M$_{500}$, as derived from SZ measurements, versus
Einstein radius for this sample of strong lensing selected galaxy clusters.  The solid line shows the median of the theoretical halo models
that include lensing selection.  The dashed lines enclose 68$\%$ of such models, and the dotted lines enclose 90$\%$.
The uncertainties on the masses are calculated by adding in quadrature
the statistical error on the Y measurement as determined by the Markov chain with the scatter at a given mass in the mass-observable relation.
There is an additional systematic uncertainty on the iteratively determined masses that results from the uncertainty in the normalization 
of the scaling relation used to determine the masses, 
represented in Figure \ref{mvr} by an error bar at the top of the figure.  
As expected, clusters with larger Einstein radii tend to have larger observed masses.  The data fall within the theoretically expected
distribution, although on average, clusters of a given mass have Einstein radii twice as large as expected, or alternatively the 
masses are expected to be about twice as high as estimated here.

To better quantify the level of agreement between the data and the theoretical predictions, we generate a Monte Carlo
simulation, populated by drawing masses from the masses of the clusters observed (varied by their uncertainties) and calculating the corresponding
Einstein radii according to the theoretically expected distribution.  We vary the Einstein radius measurements
by their associated uncertainty, and
we include a cut so that the Einstein radii are all greater than 12.5$\arcsec$ to account for any Malmquist bias effects.  
For each Monte Carlo realization, we calculate the difference between the Einstein radii of that realization and the theoretically expected Einstein radii. 
A sample with a median difference greater than we observe was found
for 2.8$\%$ of the realizations.  

The uncertainty in the amplitude of the scaling relation used to calculate mass from the SZ observations results in an uncertainty
in the masses that is correlated across the sample.  This affects the above calculation, such that the probability of selecting from the Monte Carlo 
simulation a sample as observed ranges from 2.4 to 4.4$\%$ when the scaling relation normalization is varied by its 1$\sigma$ 
statistical uncertainty. 

\subsection{Potential Systematic Effects}\label{systematics}

One concern in this analysis is that the properties which make a cluster a more efficient strong lens could also bias the 
mass measurement derived from the SZ effect.  \citet{hennawi} identified strong lenses in simulations 
and found that strong lensing clusters are preferentially aligned along the line of sight compared to the general halo population.
This could introduce a bias in our mass determinations when we fit a spherical model to the SZ data and then 
use a scaling relation derived from halos that are expected to be more randomly oriented.
When we simulate SZA data for projected triaxial models and fit a pressure model to these data, we find that for clusters 
for which the scale radius is smaller than 
the largest scales to which the SZA has sensitivity (all strong lensing clusters except for Abell 1689), the spherically integrated
Y$_{500}$ calculated 
is higher for an elongated halo aligned along the line of sight than for the same halo oriented in the plane of the sky.
Applying the scaling relation, which was derived for clusters that were not selected to be strong lenses, could cause us to overestimate the masses.
Another possibility is that an active merger could enhance a cluster's probability for strong lensing \citep{torri}, 
which could also inflate the pressure and lead to an overestimate in the SZ mass derived from the scaling relation.  
These systematic effects would make the discrepancy between theory and obervations worse.  

The scaling relation used to measure the mass from the SZ integrated Y parameter is based on weak lensing 
measurements of X-ray selected clusters.  Using simulations, \citet{meneghetti} found that on average X-ray mass measurements are biased 
toward low values (by $\sim 5$ to $10\%$), and weak lensing mass measurements have larger scatter (up to $\sim 20\%$) than X-ray mass measurements.  However, because we are basing the SZ 
mass measurements not on the assumption of hydrostatic equilibrium, but on a scaling relation with weak lensing, we would not expect
the same bias \citet{meneghetti} found in the hydrostatic mass estimates to affect the SZ masses.  

Another possibly important systematic could result from our use of the scaling relation based on the LoCuSS sample if the gas fraction, $f_{gas}$,
for LoCuSS clusters is significantly different than the typical gas fraction of the clusters in the strong lensing sample.  \citet{hicks} conducted
X-ray observations of a sample of 13 high redshift ($0.6 < z < 1.1$) optically selected clusters from RCS1 and calculated the hydrostatic masses
and gas fractions as well as the X-ray equivalent of the integrated Y parameter, Y$_X$.  They
found that the gas fraction at M$_{500}$ for high redshift ($0.6<z<1.0$) optically selected galaxy clusters is lower on average than for lower redshift, 
X-ray selected clusters and also lower than theoretical predictions.  This affects the normalization of the scaling relation 
between the hydrostatic mass and Y$_X$.  

To investigate what effect this systematic would have on our analysis, we convert the Y$_{sph,SZ}$ for the LoCuSS clusters to the equivalent
Y$_X$ according to the scaling relation from \citep{andersson} that allows a free slope and no evolution, and we recalculate the Y--M scaling 
relation based on the weak lensing masses, 
fixing the slope to the self-similar value \citep[which is equivalent
to the slope for the Y$_{X}$--M scaling relation from ][]{hicks}.  We compare the normalization of this scaling relation with the 
normalization of the scaling relation derived for the optically selected clusters by \citet{hicks}.  The difference in normalization
is about 0.4 in M at fixed Y, indicating that at fixed Y$_X$, the masses (M$_{500}$) of clusters in \citet{hicks} tend to be twice as high as in the
LoCuSS sample.  If the clusters in our sample have similarly low gas fractions compared to the LoCuSS sample, 
then the mass estimates we derive from the SZ scaling relation would be biased low.  Our sample has a higher redshift range ($0.18<z<0.9$) 
than the LoCuSS sample (with a single cluster, Abell 1689, in common to both samples), although the median redshift is not as high as 
in the \citet{hicks} sample.  One of the strong lensing clusters in our sample, RCS1J2319, is also in the sample studied by \citet{hicks}.  
They find that the gas fraction for this cluster is indeed low, and the X-ray derived hydrostatic M$_{500}$ is about a factor of 2 larger than 
the mass we derive from SZ.  X-ray measurements of $f_{gas}$ for clusters in this sample would best resolve this issue.

A number of clusters in our sample have mass measurements given in the literature that differ from the masses inferred by the SZ scaling relation.
For example, the weak lensing masses for Abell 1703, SDSSJ1531 and SDSSJ2111 \citep[scaling from M$_{vir}$ to M$_{500}$, with
measurements from ][]{masamunewl}, are systematically larger than the masses determined from the SZ scaling relation.  Different mass observables
are affected differently by properties such as alignment along the line of sight, and for strong lensing clusters that are preferentially aligned,
larger masses are expected to be measured by weak lensing compared to SZ masses.  Alternatively, if as discussed above the gas fractions for the
strong lensing clusters were systematically lower than for the LoCuSS clusters, the masses we infer from the scaling relation could be 
underestimated.

\subsection{Comparison with Previous Studies}

We find that the Einstein radii are twice as large as expected for clusters of a given mass as inferred by SZ scaling relations, although the 
discrepancy has a significance of just 2$\sigma$.  Previous studies have found that Einstein radii are larger than expected
by a similar factor.  \citet{broadhurst} also find that the Einstein radii for four clusters, 
two of which (Abell 1689 and Abell 1703) are also in our sample, are twice as large as expected.  They interpret the significance of this
result to be at 4$\sigma$, assuming each cluster to be an independent measurement and finding the probability that all four are discrepant with 
theoretical expectations.  Our larger sample of strong lensing clusters enables us to better model the sample selection effects, so 
although the size of the discrepancy between the observed and predicted Einstein radii is similar between the two studies, the significance
that we calculate is lower.

\citet{locusssl} construct mass models from HST imaging of 10 low redshift clusters, adding to 10 other nearby 
strong lensing clusters from the literature, all selected as massive, X-ray luminous clusters.  They find that the
Einstein radii for their clusters are a factor of $\sim$ 2 larger than predicted, but at a significance of 1.7$\sigma$, 
which they do not claim as a detection of a discrepancy.

\citet{macswl} conduct a similar study of 12 higher redshift ($z>0.5$) X-ray selected clusters, constructing strong-lensing models based
on archival HST data.  
They find that the Einstein radii of their sample disagree with theoretical predictions by a factor of $\sim$ 1.4,
even after accounting for lensing selection effects that could boost the lensing signal relative to samples selected by other methods. 
However, 8 out of the 12 clusters in their sample do not have any spectroscopically measured redshifts for any of the lensed sources,
resulting in considerable uncertainty in the lensing models for those clusters.  

While a direct comparison of these results is not straightforward because each study uses a different set of models and measures mass
at a different radius, it is interesting
to note that other studies using different methods to measure the clusters' masses also find that the measured Einstein radii tend to be 
larger than expected from theory, though sometimes at low significance.  

\section{Summary}

We measure the SZ effect for a sample of ten galaxy clusters that were selected to have strong lensing systems.  
Observations are conducted at 31 GHz using the SZA, 
and mass measurements are derived from these observations
through a scaling relation based on weak lensing masses for the LoCuSS sample of clusters.  Comparing these masses for the strong lensing selected
clusters with their Einstein radii, derived from strong lensing mass modeling, provides information about their two-dimensional 
concentrations.  

The data are modestly inconsistent
with theoretical predictions, with evidence that the strong lenses tend to have lower masses, as measured from an SZ scaling relation,
than their Einstein radii indicate according to these models.  
A Monte Carlo simulation indicates that the probability that a sample with the median difference from the theoretically expected Einstein radii 
is greater than or equal to that observed is 2.8$\%$.  However, one possible systematic that could explain the discrepancy from 
theoretical expectations is if the clusters in our sample typically have lower gas fractions than the X-ray selected LoCuSS clusters. 
More sophisticated models may be needed to describe the observations.

This study is part of an ongoing program to acquire different mass proxies for strong lenses, including weak lensing
and dynamical masses.  Combining different mass proxies, with different 
sensitivities to the projection of line of sight structure, should
enable more robust determinations of the distribution of mass in strong lensing clusters.

\begin{acknowledgements}
Support for this work is provided by NSF through award AST-0838187 and PHY-0114422 at the University of Chicago.
Support for CARMA construction was derived from the Gordon and Betty Moore Foundation, the Kenneth T. and Eileen L. Norris Foundation, the James S. McDonnell Foundation, the Associates of the California Institute of Technology, the University of Chicago, the states of California, Illinois, and Maryland, and the National Science Foundation. Ongoing CARMA development and operations are supported by the National Science Foundation under a cooperative agreement, and by the CARMA partner universities. 
SM acknowledges support from an NSF
Astronomy and Astrophysics Fellowship; CG, SM, and MS from NSF
Graduate Research Fellowships; DPM from NASA Hubble Fellowship grant HF-51259.01
Support for TM was provided by NASA through the Einstein Fellowship Program, grant PF0-110077.
\end{acknowledgements}

{\it Facilities:} \facility{CARMA}

\clearpage
\begin{table}
\begin{center}
\caption{Strong Lensing Cluster Sample}\label{sample}
\begin{tabular}{rcccccccc}
\tableline\tableline
Name & Cluster $z^1$ & r$_{500}$  & $Y_{sph} $            & M$_{500}$                      & Arc $z^1$ & $\theta_{E}$ & $\theta_{E}$ for ($z_{l}$,$z_{s}$)=(0.5,2.0) & Lensing\\
     &             & (\arcsec) & ($10^{-5}$ Mpc$^{2}$) & $\times 10^{14}$ M$_{\odot}$ &         & (\arcsec)      & (\arcsec)                                      & Ref.$^2$\\
 \tableline
Abell 1689      &       0.18 & 503 &   $24.8^{6.5}_{-4.9}$ &        $10.9^{+2.1}_{-1.9}$ &  1.100 & 42.6 &  $40.2 \pm{2.0}$ &       1\\
&        &       &       &       &       3.000 &        52.3 &  &         \\
 &       &       &       &       &       4.800 &        54.4 &  &         \\
 Abell 1703      &      0.28 & 251 &   $4.6^{1.1}_{-0.8}$ & $3.9 \pm{0.7}$ &        0.880 & 19.2 &  $32.5 \pm{1.9}$ &       2\\
&        &       &       &       &       3.380 &        31.3 &  &         \\
 Abell 963       &      0.21 & 321 &   $4.4^{1.1}_{-0.8}$ & $3.9^{+0.7}_{-0.6}$ &   0.771 & 6.7 &   $18.5^{+2.3}_{-6.2}$ &  3\\
&        &       &       &       &       1.958 &        16.5 &  &         \\
 RCS1 J2319      &      0.90 & 85 &    $1.5^{0.6}_{-0.5}$ & $2.0 \pm{0.4}$ &        3.860 & 12.9 &  $12.8^{+4.7}_{-2.0}$ &  3\\
RCS2 J0327      &       0.56 & 133 &   $2.8^{0.4}_{-0.5}$ & $2.7 \pm{0.4}$ &        1.701 & 21.8 &  $25.3 \pm{1.5}$ &       4\\
RCS2 J2327      &       0.70 & 155 &   $12.3 \pm 0.7$ &        $6.5 \pm{0.8}$ &        1.415 & 25.9 &  $34.4^{+2.8}_{-2.1}$ &  3\\
&        &       &       &       &       2.983 &        39.6 &  &         \\
 SDSS J1209$+$26 &      0.56 & 161 &   $7.3^{2.6}_{-1.9}$ & $4.9^{+1.1}_{-1.0}$ &   1.018 & 8.3 &   $23.0 \pm{1.4}$ &       3\\
&        &       &       &       &       3.949 &        27.3 &  &         \\
 SDSS J1343$+$41 &      0.42 & 134 &   $0.9^{0.4}_{-0.3}$ & $1.6 \pm{0.4}$ &        2.090 & 15.2 &  $14.1^{+1.2}_{-0.9}$ &  5\\
&        &       &       &       &       5.200 &        20.9 &  &         \\
 SDSS J1531$+$34 &      0.34 & 164 &   $1.1^{0.3}_{-0.2}$ & $1.7^{+0.4}_{-0.3}$ &   1.096 & 12.3 &  $18.0^{+2.0}_{-1.8}$ &  3\\
SDSS J2111$-$01 &       0.64 & 109 &   $2.3^{0.3}_{-0.4}$ & $2.3 \pm{0.4}$ &        2.861 & 16.2 &  $14.4^{+3.5}_{-6.0}$ &  3\\
\tableline
 \end{tabular}
 \tablecomments{$^1$ Cluster and arc redshift references are Limousin et al. (2007,2008) for Abell 1689, Abell 1703 respectively; Smith et al. (2005) for 
 Abell 963; Gilbank et al. (2008) for RCS1 J2319; Wuyts et al. (2010) for RCS2 J0327; and information on RCS2 J2327 will be published in Gladders et al. (in
 preparation).  All other cluster and arc redshifts are published in Bayliss et al. (2010b).
\\ $^2$ Lens model references are: [1] Limousin et al. (2007); [2] Limousin et al. (2008); [3] Sharon et al. in prep; [4] Wuyts et al. (2010); [5] Bayliss et al. (2010a)} \end{center}
 \end{table}

\clearpage

\end{document}